\documentclass[apex]{jjap3}

\title{Wide-Dynamic-Range Cantilever Magnetometry Using a Fiber-Optic Interferometer and its Application to High-frequency Electron Spin Resonance Spectroscopy}
\author{Hideyuki Takahashi$^{1}$\thanks{E-mail address: hide.takahashi@crystal.kobe-u.ac.jp}, Tsubasa Okamoto$^2$, Eiji Ohmichi$^{2}$, Hitoshi Ohta$^{3}$}
\inst{$^{1}$Organization for Advanced and Integrated Research , Kobe University, 1-1, Rokkodai, Nada, Kobe 657-8501, Japan\\
$^{2}$Graduate School of Science, Kobe University, 1-1 Rokkodai-cho, Nada, Kobe 657-8501, Japan\\
$^{3}$Molecular Photoscience Research Center, Kobe University, 1-1 Rokkodai-cho, Nada, Kobe 657-8501, Japan}
\abst{We present a method of broadening the dynamic range of optical interferometric detection of cantilever displacement. The key idea of this system is to use a wavelength-tunable laser source. The wavelength is subject to proportional-integral control, which is used to keep the cavity detuning constant. Under this control, the change in wavelength is proportional to the cantilever displacement. Using this technique, we can measure large displacements ($>1\ \mathrm{\mu m}$) without degrading sensitivity. We apply this technique to high-frequency electron spin resonance spectroscopy and succeed in removing an irregular background signal that arises from the constantly varying sensitivity of the interferometer.}

\begin{document}
\maketitle

Micro-cantilevers are very sensitive force sensors; they has been used not only as probes for atomic force microscopes (AFMs), but also as tools for magnetic measurements~\cite{Ohmichi2002,Rossel1996,Naughton1997,Rossel1998}. 
These cantilevers have an advantage over superconducting quantum-interference device magnetometers in that they can be used under magnetic fields larger than 10 T, and even under pulsed magnetic fields~\cite{Ohmichi2009}.
In recent years, application of this technique has been extended to quantum oscillation measurements~\cite{Ohmichi2002}, susceptibility anisotropy measurement on exotic materials~\cite{Wang2005,Okazaki2011}, and magnetic resonance force microscopy~\cite{Rugar1992,Kuehn2008}.

The cantilever displacement, $\Delta d$, can be measured using various detection methods.
The most popular of these is beam-deflection detection, which is implemented in commercial AFMs. 
On the other hand, piezoresistive~\cite{Rossel1996,Ohmichi2002}, capacitive~\cite{Naughton1997,Rossel1998}, and optical interferometric detection~\cite{Rugar1992} methods have been used for magnetic measurements
because of the limited sample space within a cryostat. 
Among these, optical interferometric detection is the most sensitive;
it can detect $\Delta d$ as a change in the interference intensity with a sensitivity of less than 10 pm. However, since the interference intensity changes against $\Delta d$　sinusoidally, this technique has difficulty in measuring $\Delta d$ values that are comparable to the laser wavelength $\lambda$ in real time. 
This is one of the reasons why piezoresistive and capacitive cantilevers have been preferred for use under high magnetic fields.

This paper presents a method for broadening the dynamic range of optical interferometric detection based on the key idea of using a wavelength-tunable laser source~\cite{Smith2009}. 
Our setup is simple and does not require any positioning mechanisms. We also demonstrate that this technique can be applied to high-frequency electron spin resonance (HFESR) spectroscopy~\cite{Ohmichi2008,HT2015}. 

\begin{figure}[tb]
	\begin{center}
		\includegraphics[width=1\hsize]{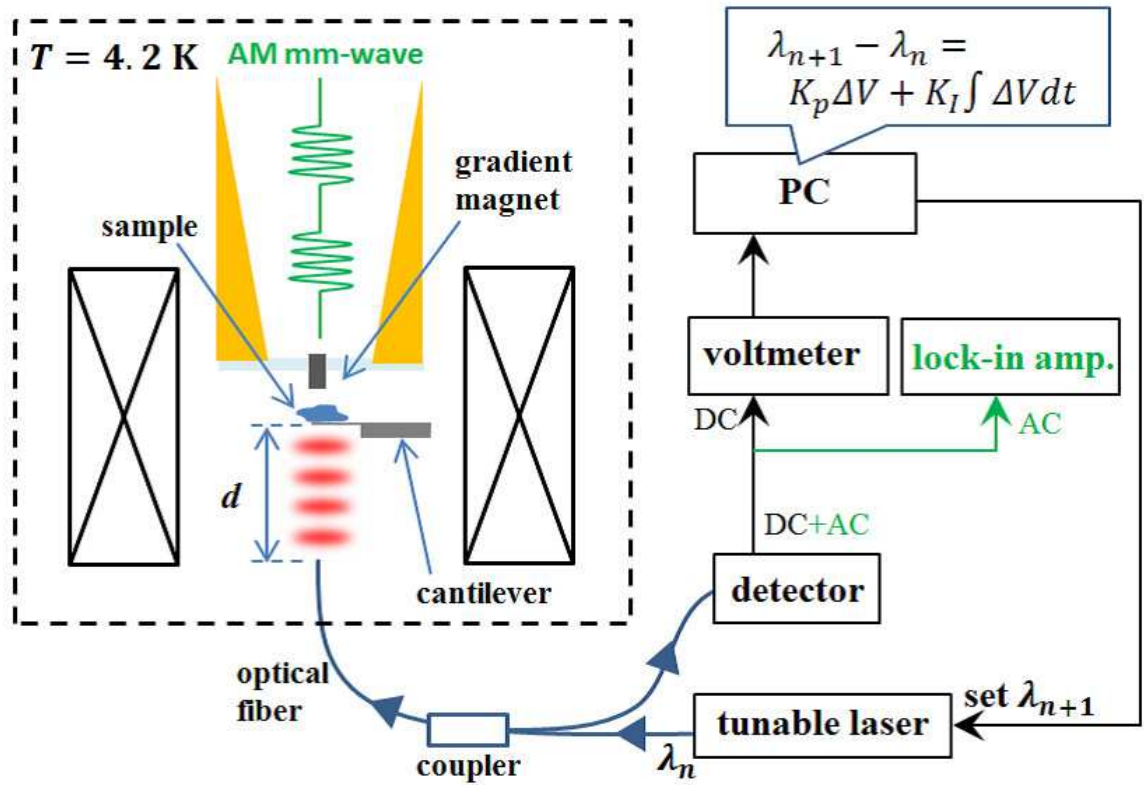}
	\end{center}
\caption{Experimental setup for cantilever magnetometry using the fiber-optic interferometer. The additional equipments for HFESR measurement are shown in green.}
\label{fig1}
\end{figure}
%faraday法の説明
Figure 1 shows the setup for the magnetization and HFESR measurements. 
It is based on the Faraday method for converting the magnetization into a magnetic gradient force by placing a sample in an inhomogeneous magnetic field. 
The sample was mounted on a commercial aluminum-coated silicon cantilever (PPP-CONTSCR by Nanosensors Inc.) with dimensions of $225\times48\times1\ \mathrm{\mu m^3}$ and a spring constant of $k=0.2\ \mathrm{N/m}$. 
A dysprosium rod was used to generate a gradient magnetic field, which was estimated to be $\partial B_z /\partial z=1000-2000\ \mathrm{T/m}$ at the sample position.
%broadband測定に対応するoversized WG 
This rod was fixed at the outlet of the circular waveguide equipped for HFESR measurements. 
The magnetic gradient force exerted upon the cantilever is given by $F\propto M_z (\partial B_z /\partial z)$, where $M_z$ is the longitudinal component of the sample magnetization.
$\Delta d$ was measured by a Fabry-Perot interferometer formed between the cantilever and the cleaved end of an optical fiber.
We used two samples, 2,2-diphenyl-1-picrylhydrazyl (DPPH) and hemin chloride (Hemin). DPPH is an organic free radical compound known to be a standard sample for ESR, while Hemin is a metal porphyrin complex with a square-planar molecular structure. The magnetic ion of Hemin is Fe(I\hspace{-.1em}I\hspace{-.1em}I) with a spin state of $S=5/2$. The sample masses, $m$, were estimated from the reduction in the eigenfrequency, $f$, of the cantilever from the unloaded value, $f_0=25\ \mathrm{kHz}$, using the relation $m=(k/4\pi^2)(f^{-2}-f_0^{-2})$.
%どちらか一方でよい？
\begin{figure}[tb]
	\begin{center}
		\includegraphics[width=0.9\hsize]{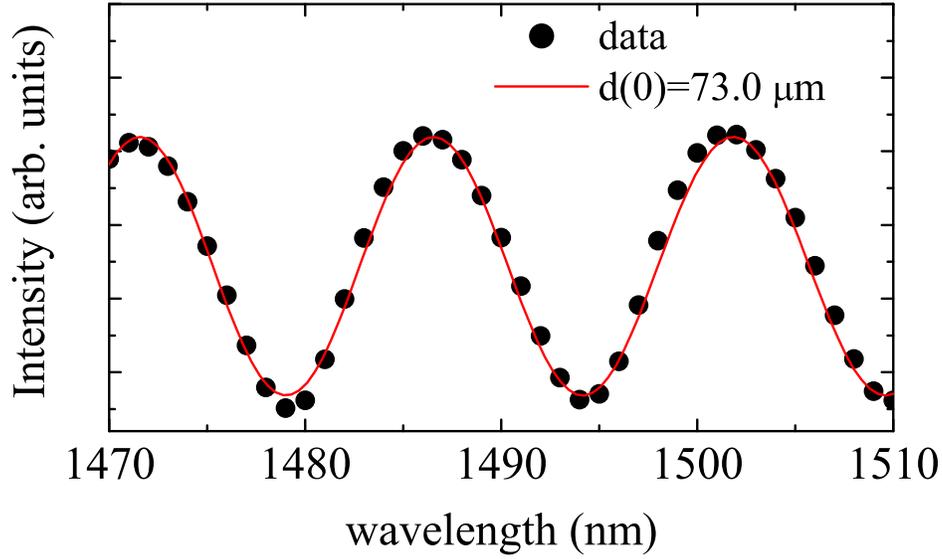}
	\end{center}
\caption{Interference intensity of the Fabry-Perot cavity formed between the cantilever and the end of the optical fiber. The red solid lines are fittings by Eq.~\ref{eq1}.}
\label{fig2}
\end{figure}

We used a wavelength-tunable laser source (Agilent Technologies, 81989A) covering the wavelength range from 1463 to 1577 nm. 
The relation between the interference signal, $V$, and cavity length, $d$, is given as
\begin{equation}\label{eq1}
\frac{V}{V_0}=1-A\cos (\frac{4\pi d}{\lambda}),
\end{equation}
where $V_0=(V_{MAX}+V_{MIN})/2$ and $A=(V_{MAX}-V_{MIN})/(V_{MAX}+V_{MIN})$. 
Figure~\ref{fig2} shows the $\lambda$ dependence of $V$ when $d=73.0\ \mathrm{\mu m}$ at a zero field. 
This corresponds to the initial condition for the measurements of Hemin in later discussions.

As shown in Fig.~\ref{fig2}, $d$ is derived from the fit for Eq.~(\ref{eq1}), or from $\lambda$ values at two adjacent maxima or minima, as $d=\lambda_m\lambda_{m+1}/2(\lambda_m-\lambda_{m+1})$~\cite{Smith2009}.
However, it is time-consuming to repeatedly measure $V(\lambda)$ during a field-swept experiment. 
Instead, we used $V$ as a feedback signal and kept the cavity detuning $d/\lambda$ constant using a software-based proportional-integral control (henceforth $\lambda$-control). 
The voltage set point $V_{\mathrm{set}}$ was determined to be $V_{\mathrm{set}}=V_{0}$.
At this voltage, $d/\lambda =1/8+n/4\ (n=0,\ 1,\ 2\cdots)$ and the maximum sensitivity was obtained.
The feedback loop time was typically 1-3 s, which was limited by the update interval of the tunable laser.

\begin{figure}[tb]
	\begin{center}
		\includegraphics[width=1\hsize]{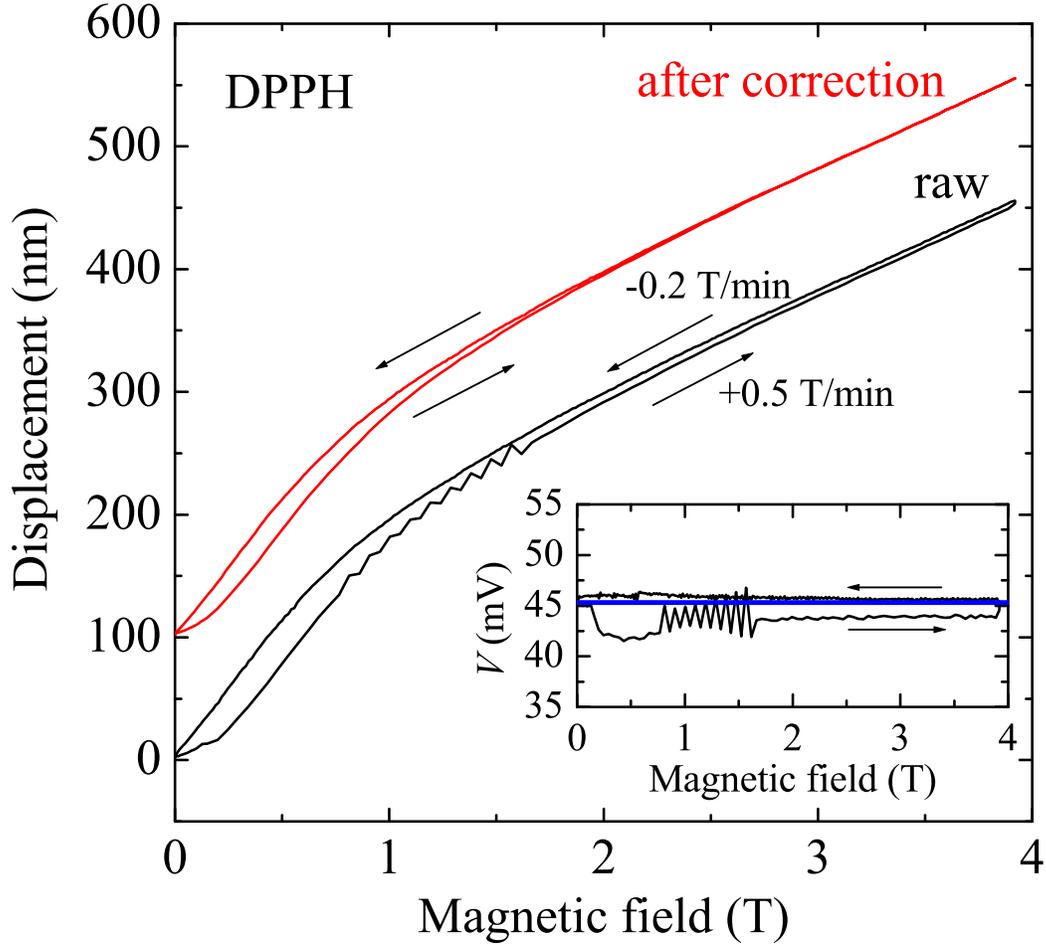}
	\end{center}
\caption{%(a) 
$\Delta d$ when DPPH was mounted. Both raw (black) and corrected data (red) are shown. The inset shows the interference intensity during the measurement. The blue line is the setpoint, $V_{\mathrm{set}}=45.3 \mathrm{mV}$. 
}
\label{fig3}
\end{figure}

Under $\lambda$-control, $d(B)$ is given as $d(B)=d(0)\lambda (B)/\lambda (0)$.
However, in practice, this equation is not sufficient for precise measurement for a number of reasons. 
First, the wavelength-tuning resolution of the laser source we used was 1 pm, which limited the resolution of a $\Delta d$ measurement to above 0.5 nm.  
Furthermore, the slow updating rate gave rise to the error from the setpoint (see the inset in Fig.~\ref{fig3}). 
Hence, a correction term is needed.
The modified relation considering these factors is given as 
\begin{equation}\label{eq2}
d(B)=\frac{d(0)}{\lambda(0)}\left(\lambda (B)+\frac{\partial \lambda(0)}{\partial V}\Big|_{V_{\mathrm{set}}}(V-V_{\mathrm{set}})\right).
\end{equation}
Here, we used the fact that $V-V_{\mathrm{set}}$ is almost linearly related to the change in $\lambda$ as long as it is much smaller than $A$.
The correction term corresponds to the first order term of the Taylor series expansion of $\lambda(0)$ around $V_{\mathrm{set}}$, whose value can be derived from the intensity curve shown in Fig.~\ref{fig2}.
Figure~\ref{fig3} shows $\Delta d$ when the $130\ \mathrm{n g}$ sample of DPPH was mounted.
To demonstrate the effectiveness of the correction, we changed the field-sweep rate between the sweep-up ($+0.5\ \mathrm{T/min}$) and -down ($-0.2\ \mathrm{T/min}$) directions. 
Although the uncorrected $\lambda(B)$ fluctuated between 0.8 T and 1.7 T in the fast sweep-up field because of the control error of $V$, its influence disappeared after correction. 
We can see the excellent agreement between the sweep-up and -down traces outside of the low-field region, where the influence of the magnetization hysteresis of dysprosium should be considered. 

\begin{figure}[tb]
	\begin{center}
		\includegraphics[width=1\hsize]{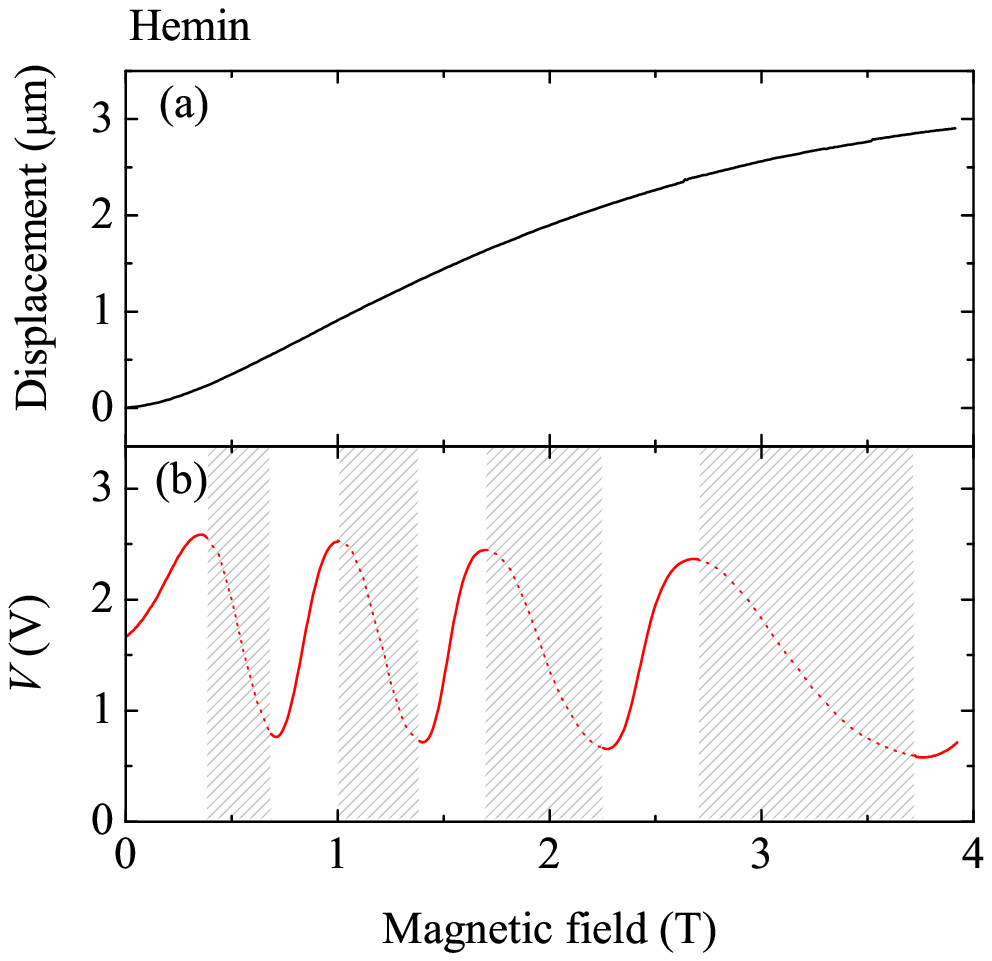}
	\end{center}
\caption{(a) $\Delta d$ when Hemin was mounted. 
(b). The red solid line shows the interference signal in the swept field obtained without $\lambda$-control. 
The shaded areas are regions where the cantilever self-oscillates.
The red dashed line in the shaded area is a guide for the eye that shows the expected behavior if self-oscillation is absent.}
\label{fig4}
\end{figure}

Figure~\ref{fig4}(a) shows $\Delta d$ when a 16 ng sample of Hemin was mounted. 
The change in $V$ without $\lambda$-control is shown in Fig.~\ref{fig4}(b).
It was found that the cantilever self-oscillates in certain magnetic field ranges.
Such oscillating behavior has been found in optomechanical micro-resonators, including micro-cantilevers,as well as in AFMs utilizing fiber-optic interferometers~\cite{Aspelmeyer2014,Metzger2004,Kim2002,Troger2010}.
It originates with the light-induced force, $F_{\mathrm{light}}$ (photothermal force or radiation pressure), from the photons inside the optical cavity.
$F_{\mathrm{light}}$ depends upon the cavity detuning factor and gives the cantilever positive or negative damping depending on its value.
In this experiment, the regions where $dV/d\lambda<0$ and $dV/d\lambda>0$ correspond to the positive- and negative- damping regions, respectively.
Since such oscillation is a nuisance for $\Delta d$ measurement, we set $\lambda(0)=1490.5\ \mathrm{nm}$ in the positive-damping region (See Fig.~\ref{fig2}) and kept the detuning constant during the measurement.

We observed a larger displacement ($\Delta d=2.90\ \mathrm{\mu m}$ at 4 T) for Hemin than that for DPPH.
The accuracy of the absolute values of $\Delta d$ was confirmed by counting the fringes of the interference intensity, $N,$ which was found in the data without $\lambda$-control.
In Fig.~\ref{fig4}(b), we can see approximately 3.8 fringes. 
Thus we obtain $\Delta d=(1/2)\lambda(0) N \simeq2.8\ \mathrm{\mu m}$, which shows a fairly good agreement with the result obtained by $\lambda$-control.
For larger $\Delta d$ measurements, the reduction in light reflection at the warped cantilever results in a decrease in $V_0$.
This would be the main source of the error in such an experiment, because cavity detuning can no longer be kept constant.
If this problem is resolved by further development, such as integrating a focusing lens, the maximum measurable displacement, $\Delta d_{\mathrm{MAX}}$, will be determined by the tunable range of $\lambda$, which is estimated to be $\Delta d_{\mathrm{MAX}}=7.8\ \mathrm{\mu m}$ for the initial conditions $\lambda (0)=1473\ \mathrm{nm}$ and $d(0)=100\ \mathrm{\mu m}$.

\begin{figure}[tb]
	\begin{center}
		\includegraphics[width=0.8\hsize]{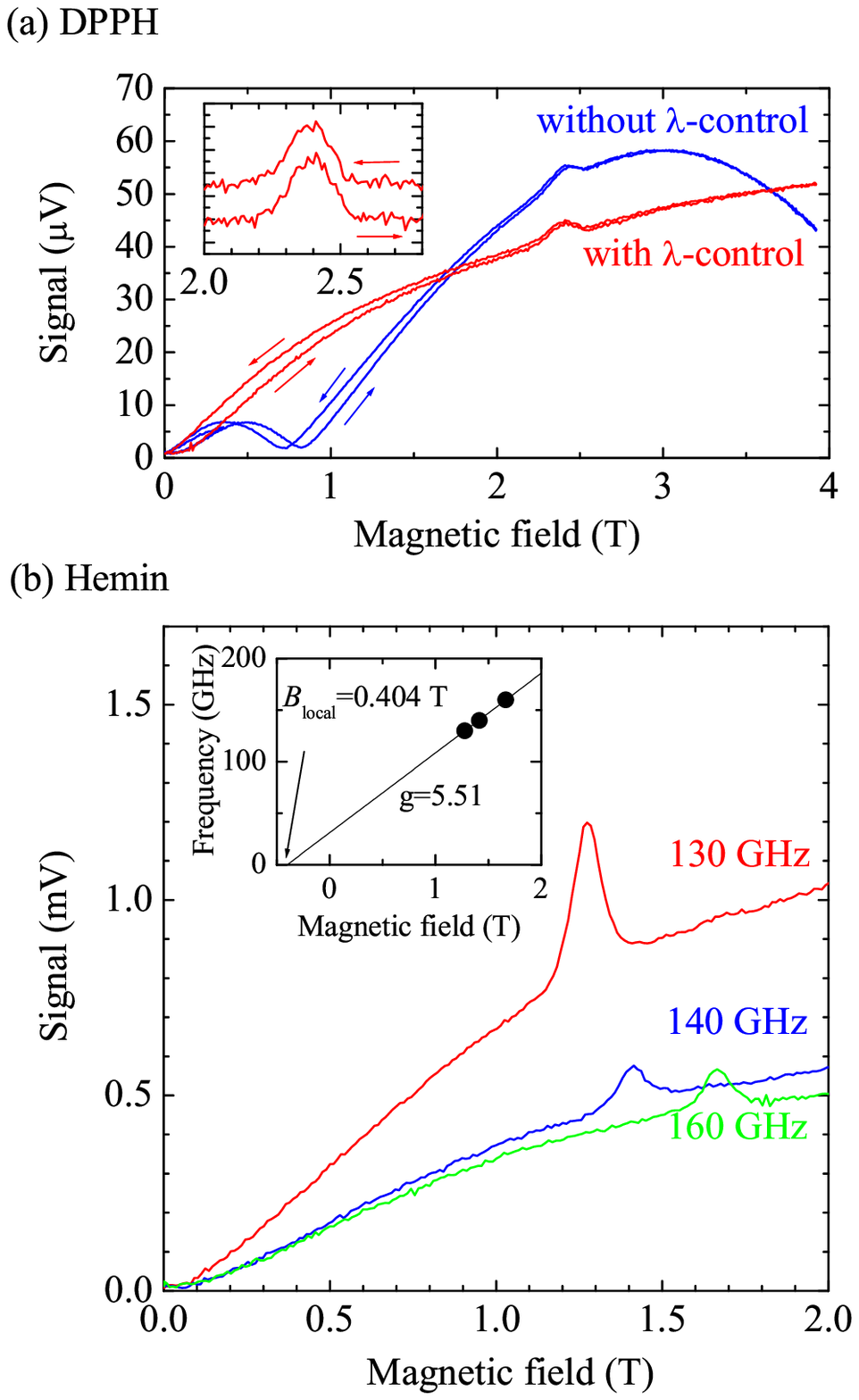}
	\end{center}
\caption{(a)HFESR spectra of DPPH at 80 GHz with (red) and without (blue) $\lambda$-control. The arrows show the directions of the field sweep. The inset shows the expanded data around the ESR signals after the smooth background signals are subtracted. (b) HFESR spectra of Hemin with $\lambda$-control. The light powers are 41, 16, and 10 mW at 130, 140, and 160 GHz, respectively. The inset shows resonant fields at each millimeter-wave frequency. The solid line is the linear fit for the data points. }
\label{fig5}
\end{figure}

$\lambda$-control is also useful for the measurements utilizing AC modulation and lock-in detection.
As an example, we demonstrate the application to HFESR spectroscopy, which is a technique for probing the local spin properties of materials.
While conventional ESR measurements are performed using the cavity perturbation or transmission methods~\cite{Poole,Motokawa1991}, ESR can also be detected as a change in magnetization because the spin-flip transition process accompanying ESR changes the occupancy of the spin states~\cite{Barco2004,Cage2005,Sakurai2011,Sakurai2012,Ohmichi2008}.
Thus far, combinations of piezoresistive cantilevers and the light-modulation technique have been very successful in enhancing spin sensitivity even in the terahertz region~\cite{Ohmichi2008,HT2015}.
However, the fact that the DC magnetization component deflects the cantilever and changes the cavity detuning has posed a problem for fiber-optic detection.
This causes a large fluctuation in the background signal and makes it difficult to observe ESR absorption.
Therefore, it is desirable to use a technique to compensate $\Delta d$ and keep the cavity detuning constant.

HFESR measurement can be performed by adding some millimeter-wave components to the setup for magnetization measurement (Fig.~\ref{fig1}).
We used an oversized circular waveguide to cover a broad frequency range between 80 GHz and 160 GHz.
The amplitude of a millimeter wave emitted from a Gunn oscillator was modulated at $f_{\mathrm{mod}}=1\ \mathrm{kHz}$ by switching the bias voltage using a TTL signal from a synthesizer.
Since a small AC signal is superimposed onto the DC signal at the detector output, it was removed before calculating $V-V_{\mathrm{set}}$ by setting the integration time of the voltmeter to be much larger than $1/f_{\mathrm{mod}}$.

Figures~\ref{fig5}(a) and (b) show the ESR spectra of DPPH and Hemin,respectively.
The intrinsic g-factor of DPPH is $g\sim 2.003$, which corresponds to a resonant field of 2.86 T at  80 GHz.
Since we used the dysprosium rod to convert magnetization into magnetic gradient force, its local field, $B_{\mathrm{local}}$, shifted the absorption peak toward a lower frequency.
It also made the internal magnetic field inhomogeneous, which resulted in broadening of the line width from its intrinsic value ($\simeq 20\ \mathrm{G}$).
For the measurement of Hemin, $\lambda$-control was indispensable for avoiding the optomechanically-induced self-oscillation that disturbs the AC modulation technique.
Thus, these data were taken in the positive-damped condition as a magnetization measurement in Fig.~\ref{fig4}(a).
The g-factor was determined by measuring ESR at multiple frequencies.
The resonant frequency-magnetic field diagram of ESR signal is shown in the inset of Fig.~\ref{fig5}(b).
The slope and intercept on the horizontal axis of the linearly fitted line corresponds to $g=5.51$ and $B_{\mathrm{local}}= 0.404\ \mathrm{T}$, respectively.
This result is consistent with a highly anisotropic g-factor ranging from $g_{\perp}\sim2$ to $g_{\parallel}=6$, where $g_{\perp}$ and $g_{\parallel}$ are the values obtained when the magnetic field is applied perpendicularly or parallel to the basal plane in the Hemin molecule, respectively~\cite{Yonetani1967}.

Without $\lambda$-control, the influence of varying sensitivity was seen in the irregular shape of the background signal (Fig~\ref{fig5}(a)).
On the other hand, the data obtained using $\lambda$-control were on the smooth off-resonant background.
The off-resonant signal had a tendency to evolve as the millimeter-wave power increased.
This indicates that it originated with the modulation of sample magnetization under periodic photo-thermal heating. When the magnetization curve is described by a Brillouin function, the temperature increase, $\Delta T$, causes the change in magnetization, $\Delta M_{\mathrm{thermal}}=(\partial M_z /\partial T)|_{4.2 \mathrm{K}}\Delta T\propto B(\partial M_z /\partial B)|_{4.2 \mathrm{K}}\Delta T$. This explains the behavior shown in Fig.~\ref{fig5}. We also confirmed that $\Delta M_{\mathrm{thermal}}$ approached zero again in the high-field limit where $\partial M_z /\partial B\rightarrow 0$. 

A similar feedback control for $\Delta d$ measurement is also achieved using piezoelectric transducers (PZTs)~\cite{Rugar1989}.
In that case, $d$ is kept constant by compensating for $\Delta d$ with extension or compression of PZT.
$\Delta d$ is obtained from the voltage applied to the electrodes.
The advantages of $\lambda$-control over the PZT method are stability and ease of use.
Although hysteresis, nonlinearity, and temperature dependence of the piezoelectric coefficient should be considered for PZT devices, the present method does not require any calibration.  
In particular, it is useful for measurements at cryogenic temperatures, where a large PZT is needed to obtain a movable range larger than $1\ \mathrm{\mu m}$ because the piezoelectric coefficient decreases to 1/5-1/10 of the value at room temperature.

In conclusion, we developed a magnetization measurement system using a micro-cantilever and a fiber-optic interferometer.
By controling $\lambda$ such that cavity detuning was kept constant, we suceeded in measuring $\Delta d$ up to $3\ \mathrm{\mu m}$ without degrading sensitivity.
We demonstrated the applicability of this method to AC modulation and  the lock-in detection technique by performing HFESR spectroscopy.
%10um程度の変位が測定できるポテンシャルがある。

This study was partly supported by a Grant-in-Aid for Young Scientists (B) (16K17749), Grant-in-Aid for Scientific Research (B) (No. 26287081), and the Asahi Glass Foundation.

\end{document}